\documentclass[aps,pre,twocolumn,groupedaddress,showpacs,floatfix]{revtex4}
\usepackage{graphicx}
\usepackage{pst-all}
\usepackage{color}
\usepackage{amsmath}
\newcommand{\comment}[1]{}

\begin{document}
\title{Random Boolean Networks}
\author{Barbara Drossel}
\affiliation{Institute of Condensed Matter Physics, Darmstadt University of Technology, Hochschulstra\ss e 6, 64289 Darmstadt, Germany }
\date{June 2007}
\begin{abstract}
 This review explains in a self-contained way the properties of
  random Boolean networks and their attractors, with a special focus
  on critical networks. Using small example networks, analytical
  calculations, phenomenological arguments, and problems to solve, the
  basic concepts are introduced and important results concerning phase
  diagrams, numbers of relevant nodes and attractor properties are
  derived

\end{abstract}

\maketitle

\section{Introduction}
\label{intro}

Random Boolean networks (RBNs) were introduced in 1969 by S. Kauffman
\cite{kau1969a,kau1969b} as a simple model for gene regulatory
networks. Each gene was represented by a node that has two possible
states, ``on'' (corresponding to a gene that is being transcribed) and
``off'' (corresponding to a gene that is not being transcribed). There
are altogether $N$ nodes, and each node receives input from $K$
randomly chosen nodes, which represent the genes that control the
considered gene. Furthermore, each node is assigned an update function
that prescribes the state of the node in the next time step, given the
state of its input nodes. This update function is chosen from the set
of all possible update functions according to some probability
distribution. Starting from some initial configuration, the states of
all nodes of the network are updated in parallel. Since configuration
space is finite and since dynamics is deterministic, the system must
eventually return to a configuration that it has had before, and from
then on it repeats the same sequence of configurations periodically:
it is on an \emph{attractor}.

S. Kauffman focussed his interest on \emph{critical} networks, which
are at the boundary between \emph{frozen} networks with only very
short attractors and \emph{chaotic} networks with attractors that may
include a finite proportion of state space.  He equated attractors
with cell types. Since each cell contains the same DNA (i.e., the same
network), cells can only differ by the pattern of gene activity. Based
on results of computer simulations for the network sizes possible at
that time, S. Kauffman found that the mean number of attractors in
critical networks with $K=2$ inputs per node increases as
$\sqrt{N}$. This finding was very satisfying, since the biological
data available at that time for various species indicated that the
number of cell types is proportional to the square root of the number
of genes. This would mean that the very simple model of RBNs with its
random wiring and its random assignment of update functions displays
the same scaling laws as the more complex reality. The concept of
universality, familar from equilibrium critical phenomena, appeared to
work also for this class of nonequilibrium systems.  Kauffman found
also that the mean length of attractors increases as $\sqrt{N}$.

Today we know that the biological data and the computer simulation
data are both incorrect. The sequencing of entire genomes in recent
years revealed that the number of genes is not proportional to the
mass of DNA (as was assumed at that time), but much smaller for
higher organisms.  The square-root law for attractor numbers and
lengths in RBNs survived until RBNs were studied with much more
powerful computers. Then it was found that for larger $N$ the apparent
square-root law does not hold any more, but that the increase with
system size is faster. The numerical work was complemented by several
beautiful analytical papers, and today we know that the attractor
number and length of $K=2$ networks increases with network size faster
than any power law. We also know that, while attractor numbers do not
obey power laws, other properties of critical RBNs do obey power laws.

It is the purpose of this review to explain in an understandable and
self-contained way the properties of RBNs and their attractors, with a
special focus on critical networks. To this aim, this review contains
examples, short calculations, phenomenological arguments, and problems
to solve. Long calculations and plots of computer simulation data were
not included and are not necessary for the understanding of the
arguments. The readers will also benefit from consulting the review
\cite{aldana}, which, while not containing the more recent findings,
covers many important topics related to Boolean networks. 

Boolean networks are used
not only to model gene regulation networks, but also neural networks,
social networks, and protein interaction networks. The structure of
all these networks is different from RBNs with their random wiring and
random assignment of update functions, and with the same number of
inputs for every node. Nevertheless, understanding RBNs is a first and
important step on our way to understanding the more complex real
networks.

\section{Model}
\label{model}

A random Boolean network is specified by its topology and its
dynamical rules. The topology is given by the nodes and the links
between these nodes. The links are directed, i.e., they have an arrow
pointing from a node to those nodes that it influences.  The dynamical
rules describe how the states of the nodes change with time. The
state of each node is ``on'' or ``off'', and it is determined by the
state of the nodes that have links to it (i.e., that are its
inputs). In the following, we first describe the topology, and then
the dynamics of RBNs.

\subsection{Topology}

\begin{figure}[htb!]
\begin{center}
\includegraphics[width=0.75\columnwidth]{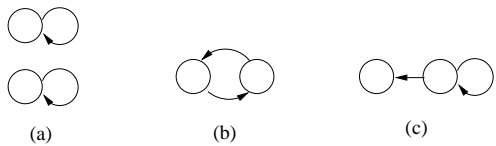}
\caption{The possible topologies for $N=2$ and $K=1$.}
\end{center}
\label{fig1}
\end{figure}

For a given number
$N$ of nodes and a given number $K$ of inputs per node, a RBN is
constructed by choosing the $K$ inputs of each node at random among
all nodes. If we construct a sufficiently large number of networks in
this way, we generate an \emph{ensemble} of networks. In this
ensemble, all possible topologies occur, but their statistical weights
are usually different. Let us consider the simplest possible example,
$N=2$ and $K=1$, shown in Figure \ref{fig1}.
There are 3 possible topologies.  Topologies (a) and (b) have each the
statistical weight 1/4 in the ensemble, since each of the links is
connected in the given way with probability 1/2.  Topology (c) has the
weight 1/2, since there are two possibilities for realizing this
topology: either of the two nodes can be the one with the self-link.

While the number of inputs of each node is fixed by the parameter $K$,
the number of outputs (i.e. of outgoing links) varies between the
nodes.  The mean number of outputs must be $K$, since there must be in
total the same number of outputs as inputs. A given node becomes the
input of each of the $N$ nodes with probability $K/N$. In the
\emph{thermodynamic limit} $N \to \infty$ the probability distribution
of the number of outputs is therefore a Poisson distribution 
\begin{equation}
P_{out}(k) = \frac {K^k}{k!} e^{-K} \, .  \end{equation}

\subsection{Update functions}

Next, let us specify the dynamics of the networks. Each node can be in
the state $\sigma_i=1$ (``on'') or in the state $\sigma_i = 0$
(``off''), where $i$ is the index of the node. The $N$ nodes of the
network can therefore together assume $2^N$ different states.  An
update function specifies the state of a node in the next time step,
given the state of its $K$ inputs at the present time step.  Since
each of the $K$ inputs of a node can be on or off, there are $M=2^K$
possible input states. The update function has to specify the new
state of a node for each of these input states.  Consequently, there
are $2^{M}$ different update functions.

\begin{table} 
\caption{The 4 update functions for nodes with 1 input. The first
column lists the 2 possible states of the input, the other
columns represent one update function each, falling into two
classes.}  
\begin{center}
\begin{tabular}{|c||c|c||c|c|}\hline In&
\multicolumn{2}{|c||}{$\mathcal{F}$}&
\multicolumn{2}{|c|}{$\mathcal{R}$}\\\hline
0&1&0&0&1\\
1&1&0&1&0\\
\hline \end{tabular} \end{center}
\label{tab1} \end{table} 

Table \ref{tab1} lists the 4 possible update functions for $K=1$. The
first two
functions are constant, or ``frozen'', i.e. the state of the node is
independent of its inputs. The other two functions change whenever an
input changes, i.e., they are \emph{reversible}. The third function is
the ``copy'' function, the fourth is the ``invert'' function. 

Table \ref{tab2} lists the 16 possible update functions for
$K=2$. There are again two constant and two reversible
functions. Furthermore, there are \emph{canalizing} functions. A
function is canalyzing if at least for one value of one of its inputs
the output is fixed, irrespective of the values of the other
inputs. The first class of canalyzing functions do not depend at all
on one of the two inputs. They simply copy or invert the value of one
of the inputs. In Table \ref{tab2}, these are the ${\mathcal{C}}_1$
functions. The second class of canalyzing functions has three times a
1 or three times a 0 in its output (the ${\mathcal{C}}_2$
functions). For each of the two inputs there exists one value that
fixes the output irrespective of the other input. In fact, constant
functions can also be considered as canalyzing functions, because the
output is fixed for any value of the inputs. 

\begin{table}
\caption{The 16 update functions for nodes with 2 inputs. The first
column lists the 4 possible states of the two inputs, the other
columns represent one update function each, falling into four
classes.}  
\begin{center}
\begin{tabular}{|c||c|c||c|c|c|c||c|c|c|c|c|c|c|c||c|c|}\hline In&
\multicolumn{2}{|c||}{$\mathcal{F}$}&
\multicolumn{4}{|c||}{${\mathcal{C}}_1$}&
\multicolumn{8}{|c||}{${\mathcal{C}}_2$}&
\multicolumn{2}{|c|}{$\mathcal{R}$}\\\hline
00&1&0&0&1&0&1&1&0&0&0&0&1&1&1&1&0\\
01&1&0&0&1&1&0&0&1&0&0&1&0&1&1&0&1\\
10&1&0&1&0&0&1&0&0&1&0&1&1&0&1&0&1\\
11&1&0&1&0&1&0&0&0&0&1&1&1&1&0&1&0\\\hline \end{tabular} \end{center}
\label{tab2} \end{table} 

Each node in the network is assigned an update function by randomly
choosing the function from all possible functions with $K$ inputs
according to some probability distribution. The simplest probability
distribution is a constant one. For $K=2$ networks, each function is
then chosen with probability 1/16. 
In the previous section, we have introduced the concept an ensemble of
networks. If we are only interested in topology, an ensemble is
defined by the values of $N$ and $K$. When we want to study network
dynamics, we have to assign update functions to each network, and the
ensemble needs to be specified by also indicating which probability
distribution of the update functions shall be used. If all 4 update
functions are allowed, there are 36 different networks in the ensemble
shown in Figure \ref{fig1}. For topologies (a) and (b), there are 10
different possiblities to assign update functions, for topology (c)
there are 16 different possibilities. The determination of the
statistical weight of each of the 36 networks for the case that every
update function is chosen with the same probability is left to the
reader....

In the following we list several frequently used probability distributions for the update functions. Throughout this article, we will refer to these different ``update rules''.
\begin{enumerate}
\item Biased functions: A function with $n$ times the output value 1
  and $M-n$ times the output value 0 is assigned a probability $p^n
  (1-p)^{M-n}$.  Then the two frozen functions in table \ref{tab2}
  have the probabilities $p^4$ and $(1-p)^4$, each of the ${\mathcal{C}}_1$
  functions and of the reversible functions has the probability $p^2(1-p)^2$,
  and the ${\mathcal{C}}_2$ functions have the probabilities $p(1-p)^3$ and
  $p^3(1-p)$.  For the special case $p=1/2$, all functions have the
  same probability 1/16.
\item Weighted classes: All functions in the same class are assigned the same
  probability. $K=1$ networks are most interesting if the two
  reversible functions occur with probability 1/2 each, and the two
  constant functions with probability 0. In general $K=1$ networks, we
  denote the weight of the constant functions with $\delta$.  An
  ensemble of $K=2$ networks is  specified by the four parameters
  $\alpha$, $\beta$, $\gamma$, and $\delta$ for the weight of ${\mathcal{C}}_1$,
  reversible, ${\mathcal{C}}_2$ and frozen functions. The sum of the
  four weights must be 1, i.e., $1=\alpha+\beta+\gamma+\delta $.
\item Only canalyzing functions are chosen, often including the
  constant functions. This is motivated by the finding that gene regulation networks appear to have many canalyzing functions and by considerations that canalyzing functions are biologically meaningful \cite{harris02,kauffman04}.
Several authors create canalyzing networks using
  three parameters \cite{moreira-amaral}. One input of the node is
  chosen at random to be a canalyzing input. The first parameter,
  $\eta$, is the probability that this input is canalyzing if its
  value is 1. The second parameter, $r$, is the probability that the
  output is 1 if the input is on its canalyzing value. 
The third parameter,
  $p$, assigns update functions for the $K-1$ other inputs according
  to rule 1 (biased functions), for the case that the canalyzing input
  is not on its canalyzing value. (This notation is not uniform
  throughout literature. For instance, in \cite{moreira-amaral}, the
  second and third parameter are named  $\rho_1$ and $\rho_2$.)
\item Only threshold functions are chosen, i.e. the update rule is
\begin{equation}
\sigma_i(t+1) =  \left\{ 
\begin{array}{l}
1 \mbox{ if } \sum_{j=1}^N \left(c_{ij}(2\sigma_j-1)+h\right) \ge 0\\
0 \mbox{ else}
 \end{array}
\right.  \end{equation} The couplings $c_{ij}$ are zero if node $i$ receives no
input from node $j$, and they are $\pm 1$ with equal probability if
node $j$ is an input to node $i$. Negative couplings are
\emph{inhibitory}, positive couplings are \emph{excitatory}. The
parameter $h$ is the \emph{threshold}. Threshold networks are inspired
by neural networks, but they are also used in some models for gene
regulation networks \cite{li:yeast,rohlf:criticality,bornholdt:robustness}.  
\item All nodes are assigned the same function. The network is then a
  cellular automaton with random wiring. 
\end{enumerate}

\subsection{Dynamics}

Throughout this paper, we only consider the case of parallel
update. All nodes are updated at the same time according to the state
of their inputs and to their update function. Starting from some
initial state, the network performs a trajectory in state space and
eventually arrives on an \emph{attractor}, where the same sequence of
states is periodically repeated. Since the update rule is
deterministic, the same state must always be followed by the same next
state. If we represent the network states by points in the
$2^N$-dimensional state space, each of these points has exactly one
``output'', which is the successor state. We thus obtain a graph in
state space.

The \emph{size} or \emph{length  of an attractor} is the number of different states on
the attractor. 
The \emph{basin of attraction} of an attractor is the set of all
states that eventually end up on this attractor, including the
attractor states themselves. The size of the basin of attraction is
the number of states belonging to it. The graph of states in state
space consists of unconnected components, each of them being a basin
of attraction and containing an attractor, which is a loop in state
space. The \emph{transient} states are those that do not lie on an
attractor. They are on  \emph{trees} leading to the attractors. 

Let us illustrate these concepts by studying the small $K=1$ network shown in Figure 2.2, which consists of 4 nodes:
\begin{figure}[htb!]
\begin{center}
\includegraphics[width=0.3\columnwidth]{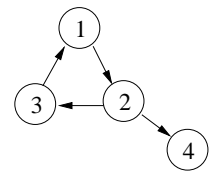}
\caption{A small network with $K=1$ input per node.}
\end{center}
\label{fig2}
\end{figure}

\begin{figure}[htb!]
\begin{center}
\includegraphics[width=0.62\columnwidth]{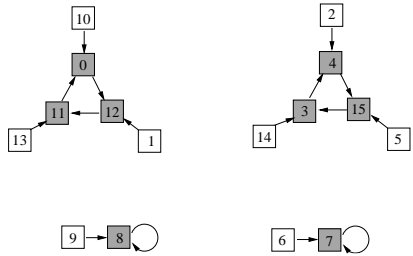}
\caption{The state space of the network shown in Figure 2.2, if the functions copy, copy, invert, invert are assigned to the four nodes. The numbers in the squares represent states, and arrows indicate the successor of each state. States on attractors are shaded.}
\end{center}
\label{fig3}
\end{figure}

If we assign to the nodes 1,2,3,4 the functions invert, invert, copy, copy, an initial state 1111 evolves in the following way:
$$ 1111 \to 0011 \to 0100 \to 1111 $$
This is an attractor of period 3. If we interpret the bit sequence characterizing the state of the network as a number in binary notation, the sequence of states can also be written as
$$15 \to 3 \to 4 \to 15 $$
The entire state space is shown in Figure \ref{fig3}.

There are 4 attractors, two of which are fixed points (i.e., attractors of length 1). The sizes of the basins of attraction of the 4 attractors are 6,6,2,2. If the function of node 1 is a constant function, fixing the value of the node at 1, the state of this node fixes the rest of the network, and there is only one attractor, which is a fixed point. Its basin of attraction is of size 16. If the functions of the other nodes remain unchanged, the state space then looks as shown in Figure 2.4.
\begin{figure}[htb!]
\begin{center}
\includegraphics[width=0.3\columnwidth]{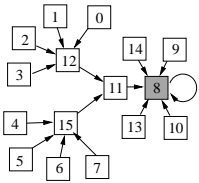}
\caption{ The state space of the network shown in Figure 2.2, if the functions 1, copy, invert, invert are assigned to the four nodes. }
\end{center}
\label{fig4}
\end{figure}

Before we continue, we have to make the definition of attractor more
precise: as the name says, an attractor ``attracts'' states to
itself. A periodic sequence of states (which we also call
\emph{cycle}) is an attractor if there are states outside the
attractor that lead to it.  However, some networks contain cycles that
cannot be reached from any state that is not part of it.
For instance, if we removed node 4 from the network shown in Figure
2.2, the state space would only contain the cycles shown in Figure
\ref{fig3}, and not the 8 states leading to the cycles. In the
following, we will use the word ``cycle'' whenever we cannot be
confident that the cycle is an attractor.

\subsection{Applications}

Let us now make use of the definitions and concepts introduced in this
section in order to derive some results concerning cycles in state
space. First, we prove that in an ensemble of networks with update
rule 1 (biased functions) or rule 2 (weighted classes), there is on an
average exactly one fixed point per network.  A fixed point is a cycle
of length 1.  The proof is slightly different for rule 1 and rule
2. Let us first choose rule 2. We make use of the property that for
every update function the inverted function has the same
probability. The inverted function has all 1s in the output replaced
with 0s, and vice versa.  Let us choose a network state, and let us
determine for which fraction of networks in the ensemble this state is
a fixed point. We choose a network at random, prepare it in the chosen
state, and perform one update step. The probability that node 1
remains in the same state after the update, is 1/2, because a network
with the inverted function at node 1 occurs equally often. The same
holds for all other nodes, so that the chosen state is a fixed point
of a given network with probability $2^{-N}$. This means that each of
the $2^N$ states is a fixed point in the proportion $2^{-N}$ of all
networks, and therefore the mean number of fixed points per network is
1. We will see later that fixed points may be highly clustered: a
small proportion of all networks may have many fixed points, while the
majority of networks have no fixed point.

Next, we consider rule 1. We make now use of the property that for
every update function a function with any permutation of the input
states has the same probability. This means that networks in which
state A leads to state B after one update, and networks in which
another state C leads to state B after one update, occur equally often
in the ensemble.  Let us choose a network state with $n$ 1s and $N-n$
0s. The average number of states in a network leading to this state
after one update is $2^N p^n (1-p)^{N-n}$. Now, every state leads
equally often to this state, and therefore this state is a fixed point
in the proportion $p^n (1-p)^{N-n}$ of all networks. Summation over
all states gives the mean number of fixed points per network, which is
1.

Finally, we derive a general expression for the mean number of cycles
of length $L$ in networks with $K=2$ inputs per node. The
generalization to other values of $K$ is straightforward.  Let
$\langle C_L\rangle_N$ denote the mean number of cycles in state space
of length $L$, averaged over the ensemble of networks of size $N$. On
a cycle of length $L$, the state of each node goes through a sequence
of 1s and 0s of period $L$. Let us number the $2^{L}$ possible
sequences of period $L$ of the state of a node by the index $j$,
ranging from 0 to $m= 2^{L}-1$. Let $n_j$ denote the number of nodes that have the sequence $j$
on a cycle of length $L$, and $(P_L)_{l,k}^j$ the probability that a
node that has the input sequences $l$ and $k$ generates the output
sequence $j$. This probability depends on the probability distribution
of update functions. Then
\begin{equation}
\langle C_L\rangle_N = \frac 1 L \sum_{\{n_j\}} \frac{N!}{n_0!\dots n_{m}!}\prod_j\left(\sum_{l,k} \frac{n_{l}n_{k}}{N^2}(P_L)_{l,k}^j\right)^{n_j}\, . \label{st}
\end{equation}
 The factor $1/L$ occurs because any of the $L$ states on the cycle
could be the starting point. 
The sum is over all
possibilities to choose the values $\{n_j\}$ such that $\sum_j n_j =
N$. The factor after the sum is the number of different ways in which
the nodes can be divided into groups of the sizes
$n_0,n_1,n_2,\dots,n_m$.
 The product is the probability that
each node with a sequence $j$ is connected to nodes with the sequences
$l$ and $k$ and has an update function that yields the output sequence
$j$ for the input sequences $l$ and $k$. 
This formula was first given in the beautiful paper by Samuelsson and
Troein \cite{samuelssontroein}. 

We conclude this section with a picture of the state space of a
network consisting of 10 nodes. 
\begin{figure}[htb!]
\begin{center}
\vskip 1.3cm
\includegraphics[width=0.5\columnwidth]{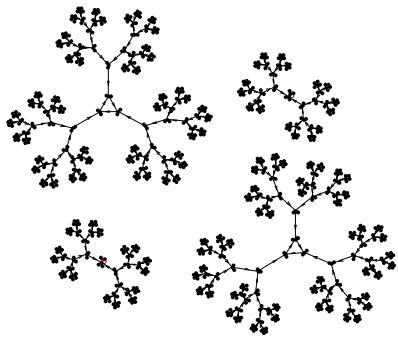}
\caption{ The state space of a network with 10 nodes}
\end{center}
\label{fig4a}
\end{figure}

\subsection{Problems}
\begin{enumerate}
\item Show that the fraction $3/32$ of all networks in the ensemble with $N=4$ and $K=1$ have the topology shown in Figure 2.2. 
\item Show that the fraction $3^3/2^{10}$ of all networks with the
  topology shown in Figure 2.2 have the state space topology shown in
  Figure 2.3, if the distribution of update functions is given by rule 1 with $p=1/4$.
\item Which functions in Table \ref{tab2} correspond to the threshold functions in networks with $K=2$, if we set $h=0$ ? 
\item Consider again $K=2$ networks, and choose the update rules 3
(canalyzing functions), which are characterized by the parameters
$\eta$, $r$, and $p$. Express the weight of each function in Table
\ref{tab2} in terms of $\eta$, $r$, and $p$.
\item Using Equation (\ref{st}), show that in an ensemble of networks
with update rule 1 or 2, there is on an average exactly one fixed
point per network.
\end{enumerate}

\section{Annealed approximation and phase diagrams}
\label{annealed}

The \emph{annealed approximation}, which is due to Derrida and Pomeau \cite{derrida86},
is a useful tool to calculate certain network properties. It is a
mean-field theory, which neglects possible correlations between
nodes. The first assumption of the annealed approximation is that the
network is infinitely large. This means that fluctuations of global
quantities are negligible. The second assumption of the annealed
approximation is that the inputs of each node can be assigned at every
time step anew. 
The following quantities can be evaluated by
the annealed approximation:
\begin{enumerate}
\item The time evolution of the proportion of 1s and 0s.
\item The time evolution of the Hamming distance between the states of
  two identical networks. 
\item The statistics of small perturbations.
\end{enumerate}
We will discuss these in the following in the order given in this
list. One of the main results of these calculations will be the phase
diagram, which indicates for which parameter values the networks are
frozen, critical or chaotic.

\subsection{The time evolution of the proportion of 1s and 0s}

Let $b_t$ denote the number of nodes in state 1, divided by $N$. The
proportion of nodes in state 0 is then $1-b_t$. We want to calculate
$b_{t+1}$ as function of $b_t$ within the annealed approximation.
Since the $K$ inputs of each node are newly assigned at each time step,
the probability that $m$ inputs of a node are in state 1 and the other
inputs in state 0 is $b_t^m (1-b_t)^{K-m}$. Since we consider an
infinitely large network, this probability is identical to the
proportion of nodes that have $m$ inputs in state 1.

 Let $p_m$ be the probability that the output value of a node with $m$
inputs in state 1 is 1. Then we have
\begin{equation}
b_{t+1} = \sum_{m=0}^K {K \choose m} p_m b_t^m (1-b_t)^{K-m} \, .
\label{bt}
\end{equation}

If $p_m$ is independent of $m$, the right-hand side is identical to
$p_m$, and $b_t$ reaches after one time step its stationary value,
which is the fixed point of Equation (\ref{bt}). Among the
above-listed update rules, this happens for rules 1 (biased functions)
and 2 (weighted classes ) and 4 (threshold functions). For
rule 1, we have $p_m=1/2$, since the output values 0 and 1 occur with
equal probability within each class of update functions. For rule 2,
we have $p_m = p$ by definition. For rule 4, the value of $p_m$ is
independent of $m$ because the value of $c_{ij}$ is 1 and $-1$ with
equal probability, making each term $c_{ij}(2\sigma_{j}-1)$ to $+1$
and $-1$ with equal probability. Therefore $p_m$ is identical to the
probability that the sum of $K$ random numbers, each of which is $+1$
or $-1$ with probability 1/2, is at least as large as $-h$,
$$ p_m = \left(\frac 1 2 \right)^K\sum\limits_{l\ge (K-h)/2} {K \choose l} \, .
$$ Here, $l$ is the number of $+1$s, and $K-l$ the number of $-1$s.

For rule 3 (canalyzing functions) we get \cite{moreira-amaral}
\begin{eqnarray}
b_{t+1} &=& b_t\eta r + (1-b_t)(1-\eta)r \nonumber \\ &&
+ b_t(1-\eta)p + (1-b_t)\eta p\nonumber\\
&=& r + \eta(p-r) + b_t(p-r)(1-2\eta)\, . \label{mapcan}
\end{eqnarray}
The first two terms are the probability that the canalyzing input is
on its canalyzing value, and that the output is then 1. The second two
terms are the probability that the canalyzing input is not on its
canalyzing value, and that the output is then 1.  This is a
one-dimensional map. The only fixed point of this map is
$$b^* = \frac{r + \eta(p-r)}{1-(p-r)(1-2\eta)}\, .$$ Since the
absolute value of the slope of this map is smaller than 1 everywhere,
every iteration (\ref{mapcan}) will bring the value of $b_t$ closer to
this fixed point.

There exist also update rules where the fixed points are unstable and
where periodic oscillations or chaos occur. This occurs particularly
easily when all nodes are assigned the same function (rule 5). For
instance, if all nodes are assigned the last one of the canalyzing
functions occurring in the table of update functions \ref{tab2}, we
have the map \begin{equation} b_{t+1} = 1-b_t^2\, . \label{thresh} \end{equation} The fixed
point $$b^* = \frac{-1+\sqrt{5}}{2}$$ is unstable, since the slope of
the map is $(1-\sqrt{5})$ at this fixed point, i.e., it has an
absolute value larger than 1. The iteration ({\ref{thresh}) moves
$b_t$ away from this fixed point, and eventually the network
oscillates between all nodes being 1 and all nodes being 0.

A map that allows for
oscillations with larger period and for chaos is obtained for the
update rule that the output is 1 only if all inputs are equal. This
map is defined for general values of $K$ and is given by
 \begin{equation} b_{t+1} = b_t^K + (1-b_t)^K\, . \label{chaoticmap} \end{equation} Let us consider $K$
as a continous parameter. When it is increased, starting
at 1, the map first has a stable fixed point and then shows a
period-doubling cascade and the Feigenbaum route to
chaos shown in Figure \ref{chaos} \cite{andreacut01}.
\begin{figure}[htb!]
\begin{center}
\vskip 0.75cm
\includegraphics[width=0.7\columnwidth]{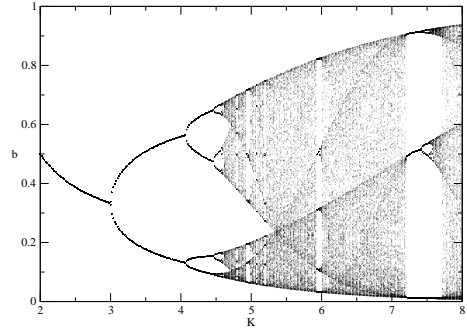}
\caption{ The values of $b_t$ that still occur after the transient time for the map (3.5), as function of $K$.  }
\end{center}
\label{chaos}
\end{figure}

All these results for $b_t$ were derived within the annealed
approximation, but they are generally believed to apply also to the
original networks with fixed connectivity patterns, if the
thermodynamic limit is taken. If this is correct, the following three
statements are also correct:
\begin{itemize}
\item
All (apart from a vanishing proportion of) initial states with a given value of $b_0$ undergo the same trajectory $b_t$ with time. 
\item This trajectory is the same for all networks (apart from a vanishing proportion).
\item When time is so large that the dynamics have reached an
attractor, the map $b_{t+1} (b_t)$ is the same as in the initial stage
for those values of $b$ that can occur on the attractors.
\end{itemize}
These assumptions appear plausible, since the paths through which a
node can affect its own input nodes are infinitely long in a randomly
wired, infinitely large network. Therefore we do not expect
correlations between the update function assigned to a node and the
states of its input nodes. Neither do we expect a correlation between
the function $b_{t+1}(b_t)$ and the question of whether a state is on an
attractor.

\subsection{The time evolution of the Hamming distance}

With the help of the Hamming distance, one can distinguish between a
frozen and a chaotic phase for RBNs. We make an identical copy of each
network in the ensemble, and we prepare the two copies of a network in
different initial states. The Hamming distance between the two
networks is defined as the number of nodes that are in a different
state. For the following, it is more convenient to use the
\emph{normalized Hamming distance}, which is the Hamming distance
divided by $N$, i.e., the proportion of nodes that are in a
different state,
\begin{equation}
h_t = \frac 1 {N} \sum_{i=1}^N \left(\sigma_i^{(1)} - \sigma_i^{(2)}\right)^2\, .\label{hdef}
\end{equation}

If $h_t$ is very small, the probability that more than one input of a
node differ in the two copies, can be neglected, and the change of
$h_t$ during one time step is given by
\begin{equation}
h_{t+1} =  \lambda h_t\, , \label{hupdate}
\end{equation}
where $\lambda$ is called the \emph{sensitivity} \cite{shmu04}.
It
is $K$ times the probability that the output of a node
changes when one of its inputs changes. 

For the first four update rules listed in Section 2.2, the value of
$\lambda$ is
\begin{eqnarray}
\lambda &=& 2Kp(1-p) \quad \hbox{(biased functions)}\nonumber\\
\lambda &=& 1- \delta \quad \hbox{(weighted classes, $K=1$)}\nonumber\\
\lambda &=& \alpha + 2\beta + \gamma =
1+\beta- \delta \quad \hbox{(weighted classes, $K=2$)}\nonumber\\
\lambda & =&
r(1-p)+(1-r)p+(K-1)(\eta(1-b_t) \nonumber\\ &&+ (1-\eta) b_t)2p(1-p) \quad\quad\quad\quad
\hbox{(canalyzing functions) }\nonumber\\
\lambda &=& K \left(\frac 1 2
\right)^{K-1} {K-1 \choose l} \quad \hbox{(threshold functions)}
\label{lambda}
\end{eqnarray}
 with $l$ in the last line being the largest integer smaller than or equal to
$(K-h)/2$.  For rule 3 (canalyzing functions), the first two terms are
the probability that the output changes when the canalyzing input is
in a different state in the two network copies; the last term is the
probability that the output changes when one of the other inputs is in
a different state in the two copies, multiplied by the number of
noncanalyzing inputs.  This is the only one out of the 4 rules where
the value of $\lambda$ depends on $b_t$ and therefore on time.

The networks are in different phases for $\lambda < 1$ and $\lambda >
1$, with the critical line at $\lambda=1$
separating the two phases. In the following, we derive the properties
of the networks in the two phases as far as possible within the
annealed approximation.

If $\lambda < 1$, the normalized Hamming distance decreases to 0. If the states
of the two copies differ initially in a small proportion of all nodes,
they become identical for all nodes, apart from possibly a limited
number of nodes, which together make a contribution 0 to the normalized Hamming
distance.  $\lambda< 1$ means also that if the two copies are
initially in identical states and the state of one node in one copy is
changed, this change propagates on an average to less than one other
node. When the two copies differ initially in a larger proportion of
their nodes, we can argue that their states also become identical
after some time: we produce a large number $Q$ of copies of the same
network and prepare their initial states such that copy number $q$ and
copy number $q+1$ (for all $q=1,...,Q$) differ only in a small
proportion of their nodes. Then the states of copy number $q$ and copy
number $q+1$ will become identical after some time, and therefore the
states of all $Q$ copies become identical (again apart from possibly a
limited number of nodes). The final state at which all copies arrive
must be a state where all nodes (apart from possibly a limited number)
become frozen at a fixed value. If the final state was an attractor
where a nonvanishing proportion of nodes go through a sequence of
states with a period larger than 1, different network copies could be in
different phases of the attractor, and the normalized Hamming distance
could not become zero. Ensembles with $\lambda < 1$ are said to be in the \emph{frozen} phase. 

All these considerations did not take into account that $\lambda$
itself may not be constant. For those update rules where $b_t$ assumes
its fixed point value after the first time step, one can apply the
reasoning of the previous paragraph starting at time step 2. For rule
3, the value $b_t$ approaches its fixed point more slowly, and
therefore the value of $\lambda$ changes over a longer time period. It
is therefore possible that the Hamming distance shows initially
another trend as during later times. Once $b_t$ has reached its fixed
point value, $\lambda$ has become constant, and if then $\lambda<1$,
the normalized Hamming distance will decrease to zero. In order to decide whether
an ensemble is in the frozen phase, one must therefore evaluate
$\lambda$ in the stationary state. 

For ensembles that have no stable stationary value of $b_t$, the above
considerations do not apply directly, since $b_t$ cycles through
different values, and so does $\lambda$. Furthermore, the two copies
may be in different phases of the cycle and will then never have a
small normalized Hamming distance. For ensembles with a finite oscillation period
$T$, one should evaluate the product of all values of $\lambda$ during
one period. If this product is smaller than 1, a small normalized Hamming
distance created in a copy of a network with a stationary oscillation,
will decrease after one period. Using a similar reasoning as before,
we conclude that then the normalized Hamming distance between any two copies of
the network will decrease to zero if they have initially the same
value of $b_t$. This means that all nodes (apart from possibly a
limited number) go through a sequence of states that has the same
period as $b_t$.

If $\lambda > 1$ when $b_t$ has reached its stationary value, the
normalized Hamming distance increases from then on with time and has a nonzero
stationary value. A change in one node propagates on an average to
more than one other node. If there is a fixed point or a short
attractor, it is unstable under many possible perturbations. There is
therefore no reason why all attractors should be short. In fact,
attractors can be very long, and the ensemble is in a phase that is
usually called \emph{chaotic}, even though this is no real chaos
because state space is finite and every trajectory becomes eventually
periodic. When $b_t$ does not become stationary but periodic, we
consider again the product of all values of $\lambda$ during one
period. If this product is larger than 1, a small normalized Hamming distance
between two copies with the same value of $b_t$ will eventually become
larger. This means that attractors can be very long and need not have
the period of $b_t$.

For $\lambda = 1$, the ensemble is at the boundary between the two
phases: it is critical. A change in one node propagates on an average
to one other node. The critical line can be obtained from
Eqs.(\ref{lambda}), leading to the phase diagram shown in 
Figure \ref{fig5}.
\begin{figure}[htb!]
\vskip 0.75cm
\begin{center}
\includegraphics[width=0.9\columnwidth]{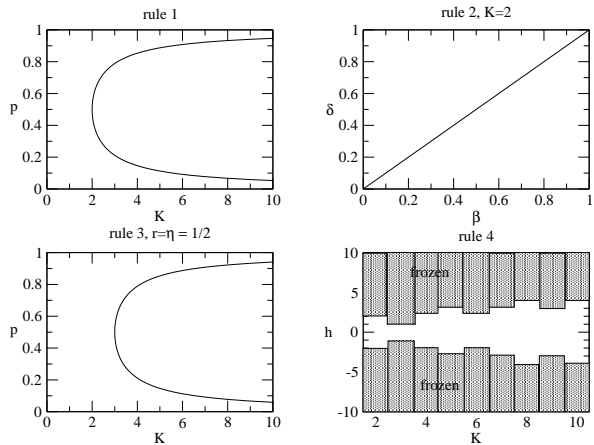}
\caption{ Phase diagram for the first 4 update rules (biased
  functions, weighted classes, canalyzing functions, threshold functions). Where there are more than 2 parameters, the remaining parameters were fixed at the values given in the respective graph titles. For theshold functions, the frozen phase is shaded. For $K=2$, the model is critical between $h=-2$ and $2$, for larger $K$, there exists no critical value, but the model is chaotic whenever it is not frozen. }
\end{center}
\label{fig5}
\end{figure}

All our results are based on a calculation for small $h_t$.
When $h_t$ is not infinitesimally small, (\ref{hupdate}) 
has the more general form
\begin{equation}
h_{t+1} =  \lambda h_t + \nu h_t^2 + \dots \, , \label{hupdategen}
\end{equation}
with the highest power of $h_t$ being $K$, but we do not make here the
effort to calculate the coefficent $\nu$ or that of a higher-order
term. We use this result only to obtain a relation between
 the stationary value of $h_t$ and the distance from the
critical line: 
In the chaotic phase, but close to the critical line (where
$\lambda$ is only slightly larger than 1), the stationary value of
$h_t$ obtained from (\ref{hupdategen}) is 
\begin{equation} h^* = (\lambda-1)/\nu\, . \label{hstat}
\end{equation}
It increases linearly with the distance from the critical line,
as long as this distance is small.

\subsection{The statistics of small perturbations in critical networks}

Now let us have a closer look at the propagation of a perturbation
that begins at one node in a critical network. Let us consider again
two identical networks, and let them be initially in the same state.
Then let us flip the state of one node in the first network. One time
step later, the nodes that receive input from this node differ in the
two systems each with probability $\lambda/K = 1/K$ (since $\lambda=1$
in a critical network). On an average, this is one node. Since the
perturbation propagates to each node in the network with probability
$(K/N) * (\lambda/K) = 1/N$, the probability distribution is a Poisson
distribution with mean value 1.  We keep track of all nodes to which
the perturbation propagates, until no new node becomes affected by
it. We denote the total number of nodes affected by the perturbation
by $s$. The size distribution of perturbations is a power law \begin{equation} n(s) \sim
s^{-3/2}\,
\label{32}\end{equation} for values of $s$ that are so small that the finite
system size is not yet felt, but large enough to see the power
law. There are many ways to derive this power law. The annealed
approximation consists in assuming that loops can be neglected, so
that the perturbation propagates at every step through new bonds and
to new nodes. In this case, there is no difference (from the point of
view of the propagating perturbation) between a network where the
connections are fixed and a network where the connections are rewired
at every time step.

We begin our calculation with one ``active'' node at time 0,
$n_a({t}=0) = 1$, which is the node that is perturbed.  At each ``time
step'' (which is different from real time!), we choose one active node
and ask to how many nodes the perturbation propagates from this node
in one step. These become active nodes, and the chosen node is now
``inactive''. We therefore have a stochastic process
$$n_a({t}+1) = n_a({t}) -1 + \xi$$ for the number of ``active'' nodes,
with $\xi$ being a random number with a Poisson distribution with mean
value 1. 
The stochastic process is finished at time ${t} =s$ when $n_a(s) =
0$. $s$ is the total number of nodes affected by the
perturbation. 

 Now we define $P_0(y,{t})$ as the probability that the stochastic
 process has arrived at $y=0$ before or at time ${t}$, if it has
 started at $n_a=y$ at time ${t}=0$. During the first step, $y$
 changes by $\Delta y = \xi-1$. If we denote the probability
 distribution of $\Delta y$ with $P(\Delta y)$, we obtain
\begin{eqnarray}
P_0(y,{t}) &=& \int d(\Delta y) P(\Delta y) P_0(y+\Delta y, {t}-1)\nonumber\\
&\simeq&  \int d(\Delta y) P(\Delta y)\nonumber\\
&& \left[P_0(y,{t}) + \Delta y \frac{\partial P_0}{\partial y} + \frac
  1 2 (\Delta y)^2 \frac{\partial^2 P_0}{\partial y^2} -\frac{\partial
    P_0}{\partial{t}}\right] \, .
\nonumber 
\end{eqnarray}
The first term on the right-hand side cancels the left-hand side. The
second term on the right-hand side is the mean value of $\Delta y$,
which is zero, times $\partial_y P_0$. Were are therefore left  with
the last two terms, which give after integration 
\begin{equation}\frac{\partial P_0}{\partial{t}} = \frac 1 2 \frac{\partial^2 P_0}{\partial y^2} \,  \label{FP}.
\end{equation}
This is a diffusion equation, and we have to apply the
initial and boundary conditions
\begin{eqnarray}
P_0(0,{t}) &=& 1\nonumber\\
P_0(y,0) &=& 0 \nonumber\\
P_0(y,\infty) &=& 1 \, .
\end{eqnarray}
Expanding $P_0$ in terms of eigenfunctions of the operator $\partial/\partial {t}$ gives the general solution
$$P_0(y,{t}) = a + by + \int d\omega\: e^{-\omega^2{t}/4}(c_\omega \sin(\omega y) + d_\omega \cos(\omega y))\, .$$
The  initial and boundary conditions fix the constants to $a=1$ and $d_\omega = 0$ and $c_\omega = -2/\pi\omega$. We therefore have 
\begin{equation}
P_0(y,{t}) = 1-\frac 2 \pi \int d\omega  \frac{\sin\omega y}{\omega}\, e^{-\omega^2{t}/4}\, ,
\end{equation}
 which becomes for $y=1$
\begin{eqnarray}
P_0(1,{t}) &=& 1-\frac 2 \pi \int d\omega \, \frac{\sin\omega }{\omega}\, e^{-\omega^2{t}/4}\nonumber\\
& \to&  1 - \mathcal{O}({t}^{-1/2}) \label{einhalb}
\end{eqnarray}
for large ${t}$. The size distribution of perturbations is obtained by taking the derivative with respect to ${t}$, leading to (\ref{32}).

Readers familar with percolation theory will notice that the spreading
of a perturbation in a critical RBN is closely related to critical
percolation on a Bethe lattice. Only the probability distribution of
the stochastic variable $\xi$ is different in this case. Since the
result depends only on the existence of the second moment of $y$, it
is not surprising that the size distribution of critical percolation
clusters on the Bethe lattice follows the same power law.

\subsection{Problems}
\begin{enumerate}
\item Explain why there is a finite critical region for $K=2$ and no critical value of $\lambda$ at all for $K>2$ for update rule 4 (Figure  \ref{fig5}). 
\item For each of the 16 update functions for $K=2$, consider an
ensemble where all nodes are assigned this function. Find the function
$b_{t+1}(b_t)$. Find all fixed points of $b$ and determine if they are
stable. If $b_t$ becomes constant for large times, determine whether
the ensemble is frozen, critical or chaotic. If $b_t$ oscillates for
large times, determine whether the normalized Hamming distance between
two identical networks that start with the same value of $b_0$, goes
to zero for large times. Interpret the result.
\item If in a frozen network only a limited number of nodes may not be frozen for large times, and if in a chaotic network a nonvanishing proportion of nodes remain nonfrozen, what do you expect in a critical network?
\end{enumerate}

\section{Networks with $K=1$}
\label{k1}

Many properties of networks with $K=1$ inputs per node can be derived
analytically. Nevertheless, these networks are nontrivial and share
many features with networks with larger values of $K$. Therefore it is
very instructive to have a closer look at $K=1$ networks. In this
review, we will not reproduce mathematically exact results that
require long calculations, as is for instance done in \cite{FK,ST2}. Instead, we will present phenomenological
arguments that reproduce correctly the main features of these networks
and that help to understand how these features result from the network
structure and update rules. We begin by studying the topology of $K=1$
networks. Then, we will investigate the dynamics on these networks in
the frozen phase and at the critical point. Finally, we will show that
the topology of $K=1$ networks can be mapped on the state space of
$K=N$ networks, which allows us to derive properties of the attractors
of $K=N$ networks, which are chaotic. 

\subsection{Topology of $K=1$ networks}

If each node has one input, the network consists of different
components, each of which has one loop and trees rooted in this loop,
as shown in Figure \ref{fig6}. 
$K=1$ networks
have the same structure as the state space pictures of other random
Boolean networks, like the ones shown in Figures 2.3 and 2.4, only the
arrows are inverted.

\begin{figure}[htb!]
\begin{center}
\includegraphics[width=0.75\columnwidth]{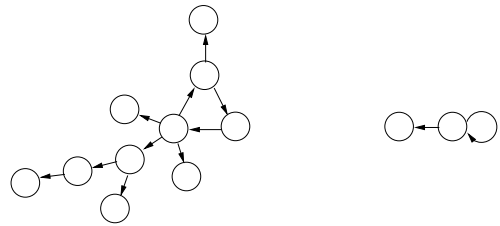}
\caption{ Example of a network with one input per node. It has
two components, the larger component has a loop of size 3 and two
trees rooted in it (one of size 1 and one of size 6), and the smaller
component has a loop of size 1 and one tree of size 1.  }
\end{center}
\label{fig6}
\end{figure}

Let us first calculate the size distribution of loops. We consider the
ensemble of all networks of size $N$. In each network of the ensemble,
each node chooses its input at random from all other nodes. The
probability that a given node is sitting on a loop of size $l$ is
therefore 
\begin{eqnarray}
P(l)&=&\left(1- \frac 1 N\right)\left(1-\frac 2 N\right)\dots \left(1-\frac{l-1}{N}\right) \frac 1 N \nonumber\\ &\simeq &  \frac{e^{-1/N}e^{-2/N}\dots
  e^{-(l-1)/N}}{N} \nonumber\\
&=& \frac{e^{-l(l-1)/2N}}{ N} \simeq \frac{e^{-l^2/2N}}{N} \, .
\end{eqnarray}
The first factor is the probability that the input to the first node
is not this node. The second factor is the probability that the input
to the second node is not the first or second node, etc. The last
factor is the probability that the input of the $l$th node is the
first node. 
The approximation in the second step becomes exact in the
thermodynamic limit $N\to \infty$ for values of $l$ that satisfy
$\lim_{N\to \infty} l/N = 0$. The approximation in the last step can be
made if $l$ is large. 

The probability that a given node is sitting on any loop is therefore
proportional to
$$\int_1^\infty P(l) dl \simeq N^{-1/2}\int_0^\infty e^{-x^2/2} dx
\propto N^{-1/2}\, .$$
This means that the total number of nodes sitting on loops is
proportional to $\sqrt{N}$. 

The mean number of nodes sitting on loops of size $l$ in a network is 
$$ NP(l) \simeq e^{-l^2/2N}\, .$$ The cutoff in loop size is
proportional to $\sqrt{N}$.  For $l \ll \sqrt{N}$, the mean number of
nodes in loops of size $l$ is 1.  This result can also be obtained by
a simple argument: The probability that a given node is sitting on a
loop of size $l$ is in the limit $N \to \infty$ simply $1/N$, since
the node almost certainly does not choose itself as input or as input
of its input etc, but in the $l$th step the first node must be chosen
as input, which happens with probability $1/N$.

The mean number of loops of size $l$ in a network is
$$ \frac{NP(l)}{l} \simeq e^{-l^2/2N}/l\, .$$ For $l \ll \sqrt{N}$, this is
simply $1/l$. Since loops are formed independently from each other in
the limit $N \to \infty$, the probability distribution of the number
of loops of size $l$ is a Poisson distribution with mean value $1/l$.

The mean number of loops per network is
$$\sum_l NP(l)/l \simeq \int _{N^{-1/2}} ^\infty \frac{e^{-x^2/2}}{x} dx \simeq
\frac 1 2 \ln N $$
for large $N$. This is identical to the mean number of components.

Next, let us consider the trees rooted in the loops. There are of the
order of $N$ nodes, which sit in $\propto \sqrt{N}$ trees, each of
which is rooted in a relevant node. This means that the average tree
size is proportional to $\sqrt{N}$.  The construction of a tree can be
described formally exactly in the same way as we described the
propagation of a perturbation in a critical network in the last
section: we begin with a node sitting in a loop. The nodes that are
not sitting in loops receive their input with equal probability from
any node in the network. Our node is therefore chosen with probability
$1/N$ by every node outside the loops as an input, and the probability
distribution of the number of outputs into the tree is a Poisson
distribution with mean value 1 (neglecting terms of the order
$N^{-1/2}$). In the same way, we find that the number of outputs of
each of the newly found tree nodes is again a Poisson distribution
with mean value 1. We iterate this process until we have identified
all nodes that are part of this tree.

The size distribution of trees is  $\sim s^{-3/2}$. The cutoff must be 
$s_{max} \sim N$ in order to be consistent with
what we know about the mean tree size and the total number of nodes in
trees: The mean tree size is 
$$\bar s \sim \int_1^{s_{max}} s s^{-3/2} ds \sim s_{max}^{1/2} \sim
\sqrt{N} \, .$$
The total number of nodes in trees is proportional to
$$\sqrt{N} \int _1^{s_{max}} s s^{-3/2} ds \sim N\, .$$

\subsection{Dynamics on $K=1$ networks}

Knowing the topology of $K=1$ networks, allows us to calculate their
dynamical properties. After a transient time, the state of the nodes
on the trees will be independent of their initial state. If a node on
a tree does not have a constant function, its state is determined by
the state of its input node at the previous time step. All nodes that
are downstream of a node with a constant function will  become
frozen. If there is no constant function in the loop and the path from
the loop to a node, the dynamics of this node is slaved to the
dynamics of the loop. 

If the weight of constant functions, $\delta$, is nonzero, 
the probability that a loop of size $l$ does not contain a frozen
function is $(1-\delta)^l$, which goes to zero when $l$ is much larger
than $1/\delta$. Therefore only loops smaller than a cutoff size can
have nontrivial dynamics. 

The number and length of the attractors of the network are determined
by the nonfrozen loops only. Once the cycles that exist on each of the
nonfrozen loops are determined, the attractors of the entire networks
can be found from combinatorial arguments.

\subsubsection{Cycles on loops}

Let us therefore focus on a loop that has no constant function. If the
number of ``invert'' functions is odd, we call the loop an odd
loop. Otherwise it is an even loop. Replacing two ``invert'' functions
with copy functions and replacing the states $\sigma_i(t)$ of the two nodes
controlled by these functions and of all nodes in between with
$1-\sigma_i(t)$, is a bijective mapping from one loop to another. In
particular, the number and length of cycles on the loop is not
changed. All odd loops can thus be mapped on loops with only one
``invert'' function, and all even loops can be mapped on loops with
only ``copy'' functions. 

We first consider even loops with only ``copy'' functions. These loops
have two fixed points, where all nodes are in the same state. If $l$
is a prime number, all other states belong to cycles of period
$l$. Any initial state occurs again after $l$ time steps. Therefore
the number of cycles on an even loop is
\begin{equation}
\frac{2^l-2}{l} + 2 \, 
\end{equation}
if $l$ is a prime number.
The numerator counts the number of states that are not fixed
points. The first term is therefore the number of cycles of length
$l$. Adding the two fixed points gives the total number of cycles. 
If $l$ is not a prime number, there exist cycles with all periods that
are a divisor of $l$. 

Next, let us consider odd loops with one ``invert'' function. After
$2l$ time steps, the loop is in its original state. If $l$ is a prime number,
there is only one cycle that has a shorter period. It is a cycle with
period 2, where at each site 0s and 1s alternate. The total number of
cycles on an odd loop with a prime number $l$ is therefore 
\begin{equation}
\frac{2^l-2}{2l} + 1 \, .
\end{equation}
If $l$ is not a prime number, there are also cycles with a period that
is twice a divisor of $l$. 

\subsubsection{$K=1$ networks in the frozen phase}

For networks with $K=1$ input per node, the parameter $\lambda$ is
\begin{equation}
\lambda = 1-\delta \, .
\end{equation}
Thefore, only networks without constant functions are critical. 
Networks with $\delta > 0$ are in the frozen phase. The mean number of
nonfrozen nodes on nonfrozen loops is given by the sum 
\begin{equation}
\sum_l (1-\delta)^l=  \frac
    {1-\delta}{\delta} \, . \label{relfrozen1}
\end{equation}
We call these loops the \emph{relevant loops}. 
We call the nodes on the relevant loops the \emph{relevant nodes}, and we denote their number with $N_{rel}$. 
The mean number of relevant loops is given by the sum
\begin{equation}
\sum_l \frac 1 l (1-\delta)^l \simeq  \ln \delta^{-1}\, ,
\end{equation}
with the last step being valid for small $\delta$.

The probability that the activity moves up the tree to the next node
is $1-\delta $ at each step.
The mean number of nonfrozen nodes on trees is therefore
\begin{equation}
 \frac
    {1-\delta}{\delta}\sum_l (1-\delta)^l= \left( \frac
    {1-\delta}{\delta}\right)^2 \, , \label{nonfrozen1}
\end{equation}
and the total mean number of nonfrozen nodes is $(1-\delta)/\delta^2$.
This is a finite number, which diverges as $\delta^{-2}$ when the critical 
point $\delta=0$ is approached.

\subsubsection{Critical $K=1$ networks}

If the proportion of constant functions $\delta$ is zero, the network is
critical, and all loops are relevant loops. There are no nodes that
are frozen on the same value on all attractors. A loop of size 1 has a
state that is constant in time, but in can take two different
values. Larger loops have also two fixed points, if they are
even. Part of the nodes in a critical $K=1$ networks are therefore
frozen on some attractors or even on all attractors, however, they can
be frozen in different states.

The network consists of $\simeq \ln N/2 $ loops, each of which has of
the order $2^l/l$ cycles of a length of the order $l$. The size of the
largest loop is of the order of $\sqrt{N}$. The number of attractors
of the network results from the number of cycles on the loops. It is
at least as large as the product of all the cycle numbers of all the
loops. If a cycle is not a fixed point, there are several options to
choose its phase, and the number of attractors of the network becomes
larger than the product of the cycle numbers. An upper bound is the
total number of states of all the loops, which is $2^{N_{rel}} \sim
e^{a\sqrt{N}}$, and a lower bound is the number of attractors on the
largest loop, which is of the order
$e^{b\sqrt{N}}/\sqrt{N}>e^{b'\sqrt{N}} $ with $b'< b < a$. From this
it follows that the mean number of attractors of critical $K=1$
networks increases exponentially with the number of relevant nodes.  A
complementary result for the number of cycles $\langle C_L \rangle $
of length $L$, which is valid for fixed $L$ in the limit $N \to
\infty$ is obtained by the following quick calculation:
\begin{eqnarray}
\langle C_L\rangle_N &\simeq &
\sum_{\{n_l\}}\prod_{l \le l_c} \left(\frac{e^{-1/l} \left(\frac 1
  {l}\right)^{n_l}}{n_l!} k_l^{n_l}\right)
\nonumber \\ &=& \sum_{\{n_l\}}\prod_{l \le l_c} \left(\frac{e^{-1/l} \left(\frac {k_l}
  {l}\right)^{n_l}}{n_l!} \right)\nonumber\\
&\simeq& \prod_{l \le l_c} e^{(k_l-1)/l} = e^{\int_1^{l_c}(k_l-1) dl/l}
  \simeq e^{(\bar k_l-1)\int_1^{l_c} dl/l}
)\nonumber\\
&\sim &  e^{(H_L-1)\ln\sqrt N} = N^{(H_L-1)/2} \, .\label{quick}
\end{eqnarray}
Here, $n_l$ is the number of loops of size $l$, $l_c$ is the cutoff in
loop size $\propto \sqrt{N}$, and $k_l$ is the number of states on a
loop of size $l$ that belong to a cycle of length $L$. This is zero
for many loops. The average over an $l$-interval of size $L$ is
identical to $H_L$, which is the number of cycles on an even loop of
size $L$. A more precise derivation of this relation, starting from
the $K=1$ version of (\ref{st}) and evaluating it by making a
saddle-point approximation, can be found in \cite{drossel05}, which is
inspired by the equivalent calculation for $K=2$ critical networks in
\cite{samuelssontroein}.

The length of
an attractor of the network is the least common multiple of the cycle
lengths of all the loops. A quick estimate gives
$$N^{a\log N}$$
since the length of the larger loops is proportional to $\sqrt{N}$,
and this has to be taken to a power which is the number of loops. A
more precise calculation \cite{ourk1paper} gives this expression,
multiplied with a factor $N^b/\log N$, which does not modify the leading
dependence on $N$.

\subsection{Dynamics on $K=N$ networks}

The topology of a $K=1$ network is identical to the topology of the
state space of a $K=N$ network, when all update functions are chosen
with the same weight. The reason is that in a $K=N$ network, the state
that succeeds a given state can be every state with the same
probability. Thus, each state has one successor, which is chosen at
random among all states. In the same way, in a $K=1$ network,
each node has one input node, which is chosen at random among all
 nodes. The state space of a $K=N$ network consists of $2^N$ nodes,
 each of which has one successor. We can now take over all results for
 the topology of $K=1$ networks and translate them into state space:

The $K=N$ networks have of the order of $\log(2^N) \propto N$
attractors. The largest attractor has a length of the order
$\sqrt{2^N} = 2^{N/2}$, and this is proportional to the total number
of states on attractors. All other states are transient states. An
attractor of length $l$ occurs with probability $1/l$ if $l \ll
2^{N/2}$.  

Clearly, $K=N$ networks, where all update functions are chosen with the same weight, are in the chaotic phase. The mean number of nodes to which a perturbation of one node propagates, is $N/2$. At each time step, half the nodes change their state, implying also that the network is not frozen.

\subsection{Application: Basins of attraction in frozen, critical and chaotic networks}

The advantage of $K=1$ networks is that they are analytically
tractable and can teach us at the same time about frozen,
chaotic and critical behavior. We will discuss in the next section to
what extent the results apply to networks with other values of
$K$. Based on our insights into $K=1$ networks, we derive now
expressions for the dependence on $N$ of the number and size of the
basins of attraction of the different attractors.

Let us first consider networks in the frozen phase. As we have seen, there is at most a limited number of small nonfrozen loops. 
Their number is independent of system size, and therefore the number of attractors is also independent of the system size. The initial state of the nodes on these nonfrozen loops determines the attractor. The initial states of all other nodes are completely irrelevant at determining the attractor. 

The size of the basin of attraction of an attractor is therefore $2^{N-N_{rel}}$, multiplied with the length of the attractor, i.e., it is $2^N$, divided by a factor that is independent of $N$. The proportion of state space belonging to a basin is therefore also independent of $N$. If we define the \emph{basin entropy} \cite{krawitz} by 
\begin{equation}
S = - \sum_a p_a \ln p_a 
\end{equation}
with $p_a$ being the fraction of state space occupied by the basin of attraction of attractor $a$, we obtain
\begin{equation} S=const \nonumber \end{equation}
for a $K=1$ network in the frozen phase. 

Next, let us consider the chaotic $K=N$ network ensemble. There are on
an average $1/l$ attractors of length $l$, with a cutoff around
$2^{N/2}$. The basin size of an attractor of length $l$ is of the
order $l 2^{N/2}$, which is $l$ times the average tree size. The basin
entropy is therefore \begin{equation} S \simeq \sum_l \frac 1 l \frac {l}{2^{N/2}}
\log \frac {l}{2^{N/2}} \simeq \int_{2^{-N/2}}^1 \log x dx = const\, .
\end{equation}

Finally, we evaluate the basin entropy for a critical $K=1$
network. There are of the order $e^{a\sqrt{N}}$ attractors with
approximately equal basin sizes, and therefore the basin entropy is
\begin{equation} S \sim \sqrt{N} \propto N_{rel}\, .  \end{equation} While frozen and chaotic
networks have a finite basin entropy, the basin entropy of critical
networks increases as the number of relevant nodes \cite{krawitz}.

\subsection{Problems}

\begin{enumerate}
\item How many cycles does an even (odd) loop of size 6 have?
\item Count the attractors of the network shown in Figure 4.1 for all four cases where loop 1 and/or loop 2 are even/odd.
\item How does the transient time (i.e. the number of time steps until
the network reaches an attractor) increase with $N$ for (a) $K=1$
networks in the frozen phase, (b) critical $K=1$ networks, (c) chaotic
$K=N$ networks?
\item Consider the subensemble of all critical $K=1$ networks that
have the same wiring, but all possible assignments of ``copy'' and
``invert'' functions. Which property determines the probability that a
network has a fixed point attractor? If it has such an attractor, how
many fixed point attractors does the network have in total? Conclude
that there is on an average one fixed point per network in this
subensemble.
\item Verify the identity $\bar k_l = H_L$ used in calculation (\ref{quick}).
\item How does the basin entropy for $K=1$ networks depend on the
parameter $\delta$ when $\delta$ becomes very small? Find an answer
without performing any calculations.
\end{enumerate}

\section{Critical Networks with $K=2$}
\label{critical}

In the previous section, we have derived many properties of frozen,
critical and chaotic networks by studying ensembles with $K=1$. Many
results are also valid for RBNs with general values of $K$. In this
section, we focus on critical $K=2$ networks. These networks, as well
as critical networks with larger values of $K$, differ in one
important respect from critical $K=1$ networks: they have a
\emph{frozen core}, consisting of nodes that are frozen on the same
value on all attractors. We have obtained this result already with the
annealed approximation: The normalized Hamming distance between two
identical networks is close to the critical point given by
(\ref{hstat}), which means that it is zero exactly at the critical
point. For $K=1$, there exists no chaotic phase and no Equation
(\ref{hstat}), and therefore the observation that all nodes may be
nonfrozen in critical $K=1$ networks is not in contradiction with the
annealed approximation.

We will first explain phenomenologically the features of
critical $K=2$ networks, and then we will derive some of these features
analytically.

\subsection{Frozen and relevant nodes}

The frozen core arises because there are constant functions that fix
the values of some nodes, which in turn lead to the fixation of the
values of some other nodes, etc. 
Let us consider Figure \ref{fig7} as an example.
\begin{figure}[htb!]
\begin{center}
\includegraphics[width=0.65\columnwidth]{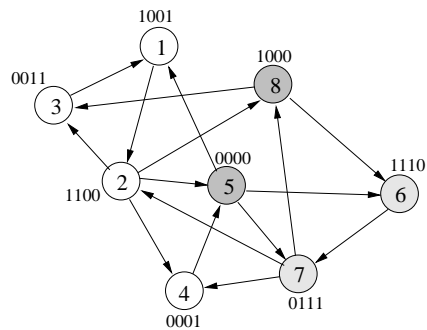}
\caption{ Example of a network with 8 nodes and $K=2$. The functions
are those of Table 2.2 (but written horizontally instead of
vertically), with the first input node being the one with the lower
number.}
\end{center}
\label{fig7}
\end{figure}
This network has the same number of constant and reversible functions,
as is required for critical networks (although this classification
only makes sense for large networks, where the thermodynamic limit
becomes visible).  Node 5 has a constant function and is therefore
frozen on the value 0 (indicated by a darker grey shade) after the
first time step. Node 6 has a canalyzing function which gives 1 as
soon as one of the inputs is 0. Therefore node 6 is frozen in state 1
(indicated by a lighter grey shade) no later than after the second
time step. Then node 7 has two frozen inputs and becomes therefore
also frozen. Its value is then 1. Node 8 has a canalyzing function
which gives 0 as soon as one of the inputs is 1, and will therefore
end up in state 0. These four nodes constitute the frozen core of this
network. At most after 4 time steps, each of these nodes assumes its
stationary value. If we remove the frozen core, we are left with a
$K=1$ network consisting of nodes 1 to 4, with ``copy'' and ``invert''
functions between these nodes. For instance, node 4 copies the state
of node 2 if node 7 is in state 1. Node 3 copies the state of node 2,
node 1 inverts the input it receives from node 3, and node 2 inverts
the input it receives from node 1.  The nodes 1,2,3 form an even loop,
and node 4 is slaved to this loop. Nodes 1,2,3 are therefore the
\emph{relevant nodes} that determine the attractors. We can conclude
that this network has 4 attractors: two fixed points and two cycles of
length 3.

There is a different mechanism by which a frozen core can arise, which
is illustrated by assigning another set of update functions to the same network, as shown in Figure 5.2.
\begin{figure}[htb!]
\begin{center}
\includegraphics[width=0.65\columnwidth]{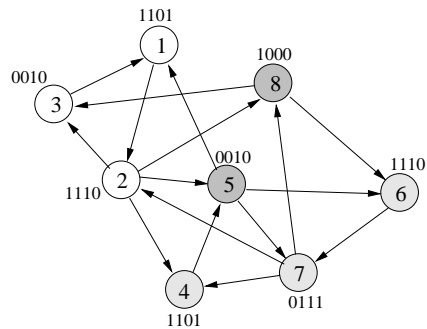}
\caption{ A network with the same topology as the previous network, but with only canalyzing functions. }
\end{center}
\label{fig8}
\end{figure}
This network contains only canalizing update functions of the type
${\mathcal{C}}_2$, and such a network could be classified as critical
if it was much larger. We begin again by fixing node 5 at value 1, and
we denote this as $5_1$. In the next time step, this node may have
changed its state, but then node 7 will be in state 1, because it is
canalyzed to this value by node 5. By continuing this consideration,
we arrive at the following chain of states:
$$5_1 \to 7_1 \to 4_1 \to 5_0\, .$$ 
This means that node 5 must eventually assume the state 0, and we continue from here by following again canalyzing connections:
$$5_0 \to 6_1 \to 7_1 \to (4_1,8_0) \to (5_0,6_1) \to (6_1,7_1)$$
$$  \to (4_1,7_1,8_0) \to (4_1,5_0,6_1,8_0)  \to (5_0,6_1,7_1) $$
$$\to (4_1,6_1,7_1,8_0) \to (4_1,5_0,6_1,7_1,8_0) \to (4_1,5_0,6_1,7_1,8_0) \, $$
From this moment on, nodes 4 to 8 are frozen. Nodes 1,2,3 form a relevant loop with the functions invert, invert, copy, just as in the previous example. 

In order to better understand how the frozen core arises in this case,
consider the loop formed by the nodes 6,7,8: This is a
\emph{self-freezing loop}. If the nodes 6,7,8 are in the states 1,1,0,
they remain forever in these states, because each node is canalyzed to
this value by the input it receives within the loop. This loop has the
same effect on the network as have nodes with constant functions. Once
this loop is frozen, nodes 4 and 5 become also frozen. One can imagine
networks where such a loop never freezes, but this becomes very
unlikely for large networks.

The networks shown in the previous two figures were designed to
display the desired properties. In general, small networks differ a
lot in the number of frozen and nonfrozen nodes, as well as in the
size and structure of their relevant component(s) and their
attractors. The specific properties particular to the frozen and
chaotic phase and to the critical line become clearly visible only for
very large networks.

A network of intermediate size is the basis of Figure 5.3,
which shows the nonfrozen part of a critical $K=2$ network with 1000
nodes. There are 100 nonfrozen nodes in this network, indicating that
the majority of nodes are frozen. Among the 100 nonfrozen nodes, only
5 nodes are relevant, and only 6 nodes have two nonfrozen inputs. The
relevant nodes are arranged in 2 relevant components. They determine
the attractors of the network, while all other nodes sit on outgoing
trees and are slaved to the dynamics of the relevant nodes.
\begin{figure}[htb!]
\begin{center}
\includegraphics[width=0.9\columnwidth]{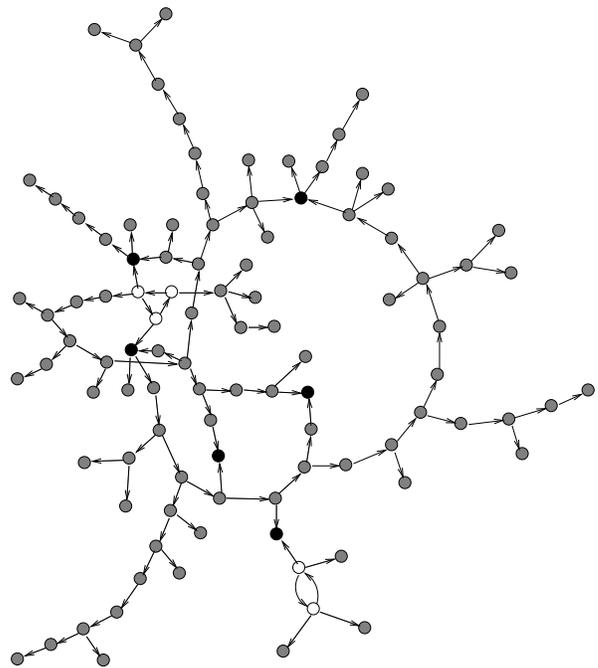}
\caption{The nonfrozen part of a $K=2$ network with 1000 nodes. Shown are the 100 nonfrozen nodes. 5 nodes are relevant (white), and 6 nodes (black) have two nonfrozen inputs.}
\end{center}
\label{fig9}
\end{figure}
This figure resembles a lot a $K=1$ network. The only difference is that there are a few nodes with two inputs. 

Analytical calculations, part of which are explained in the next section, give the following general results for critical $K=2$ networks in the thermodynamic limit $N \to \infty$:
\begin{enumerate}
\item The number
of nodes that do not belong to the frozen core, is proportional to $N^{2/3}$
for large $N$.
\item If the proportion of nodes with a constant function is nonzero, the frozen core can be determined by starting from the nodes with constant functions and following the cascade of freezing events. 
\item If the proportion of nodes with a constant function is zero (which means that the network contains only canalyzing functions), the frozen core can be determined by starting from self-freezing loops. 
\item The number of nodes that are nonfrozen and that receive 2
nonfrozen inputs is proportional to $N^{1/3}$.
\item The number of relevant nodes is proportional to $N^{1/3}$. They are connected to relevant components, which consist of loops and possibly additional links within and between the loops.
\item The number of relevant nodes that have two relevant inputs remains finite in the limit $N \to \infty$. 
\item The number of relevant components increases as $\log N^{1/3}$. 
\item The cutoff of the size of relevant components scales as $N^{1/3}$. 
\end{enumerate}
The complete list of these results is given in
\cite{kaufmanandco:ourscaling}, but part of the results can be found
in earlier papers \cite{bastolla1,bastolla2,socolar}.

\subsection{Analytical calculations} 

After this qualitative introduction to critical networks, let us
derive the main results for the scaling of the number of nonfrozen and
relevant nodes with $N$. Computer simulations of critical networks
show the true asymptotic scaling only for very larger networks with
more than 100000 nodes. For this reason, the values $2/3$ and $1/3$
for the critical exponents characterizing the number of nonfrozen and
relevant nodes has been known only since 2003.

Flyvbjerg \cite{flyvbjerg:order} was the first one to use a dynamical
process that starts from the nodes with constant update functions and
determines iteratively the frozen core. Performing a mean-field
calculation for this process, he could identify the critical point. 
We will go now beyond mean-field theory.

We consider the ensemble of all $K=2$ networks of size $N$ with update
rule 2 (weighted functions), where the weights of the
${\mathcal{C}}_1$, reversible, ${\mathcal{C}}_2$ and constant functions
are $\alpha$, $\beta$, $\gamma$ and $\delta$.  These networks are
critical for $\beta=\delta$.  We begin by assigning update functions
to all nodes and by placing these nodes according to their functions
in four containers labelled $\mathcal{F}$, $\mathcal{C}_1$,
$\mathcal{C}_2$, and $\mathcal{R}$. These containers then contain
$N_f$, $N_{c_1}$, $N_{c_2}$, and $N_r$ nodes.  We treat the nodes in
container $\mathcal{C}_1$ as nodes with only one input and with the
update functions ``copy'' or ``invert''.  As we determine the frozen
core, the contents of the containers will change with time. The
``time'' we are defining here is not the real time for the dynamics of
the system. Instead, it is the time scale for the process that we use
to determine the frozen core. One ``time step'' consists in choosing
one node from the container $\mathcal{F}$, in selecting the nodes to
which this node is an input, and in determining its effect on these
nodes. These nodes change containers accordingly. Then the frozen node
need not be considered any more and is removed from the system. The
containers now contain together one node less than before.  This means
that container $\mathcal{F}$ contains only those frozen nodes, the
effect of which on the network has not yet been evaluated. The other
containers contain those nodes that have not (yet) been identified as
frozen. The process ends when container $\mathcal{F}$ is empty (in
which case the remaining nodes are the nonfrozen nodes), or when all
the other containers are empty (in which case the entire network
freezes). The latter case means that the dynamics of the network go to the same
fixed point for all initial conditions. 

This process is put into the following equations, which describe
the changes of the container contents during one ``time step''. 
\begin{eqnarray}
\Delta N_r &=& - \frac{2N_r}{N}\nonumber \\
\Delta N_{c_2} &=& - \frac{2N_{c_2}}{N}\nonumber \\
\Delta N_{c_1} &=& \frac{2N_r}{N}+ \frac{N_{c_2}}{N}- \frac{N_{c_1}}{N} 
 \label{Delta}\\
\Delta N_f &=& -1 +  \frac{N_{c_2}}{N} + \frac{N_{c_1}}{N} + \xi\nonumber \\
\Delta N &=& -1\nonumber
\end{eqnarray}
The
terms in these equations mean the following: Each node in container
$\mathcal{R}$ chooses the selected frozen node as an input with
probability $2/N$ and becomes then a $\mathcal{C}_1$-node. This explains the first equation and the first term in the third equation. Each node
in container $\mathcal{C}_2$ chooses the selected frozen node as an
input with probability $2/N$. With probability $1/2$, it then becomes
frozen, because the frozen node is with probability $1/2$ in the state
that fixes the output of a $\mathcal{C}_2$-node. If the
$\mathcal{C}_2$-node does not become frozen, it becomes a
$\mathcal{C}_1$-node. This explains the terms proportional to $N_{c_2}$. Each node in container $\mathcal{C}_1$ chooses
the selected frozen node as an input with probability $1/N$. It then
becomes a frozen node. Finally, the $-1$ in the equation for $\Delta
N_f$ means that the chosen frozen node is removed from the system.  In
summary, the total number of nodes, $N$, decreases by one during one
time step, since we remove one node from container $\mathcal{F}$.  The
random variable $\xi$ captures the fluctuations around the mean change
$\Delta N_f$. It has zero mean and variance $(N_{c_1}+N_{c_2})/N$.
The first three equations should contain similar noise terms, but
since the final number of nodes of each class is large for large $N$,
the noise can be neglected in these equations.
We shall see below that at the end of the process most of the remaining nodes
are in container $\mathcal{C}_1$, with the proportion of nodes left in
containers $\mathcal{C}_2$ and $\mathcal{R}$ vanishing in the
thermodynamic limit. 
 Figure \ref{fig10}
illustrates the process of determining the frozen core.
\begin{figure}[htb!]
\begin{center}
\parbox{1cm}{(1)}\parbox{7cm}{\includegraphics[width=7cm]{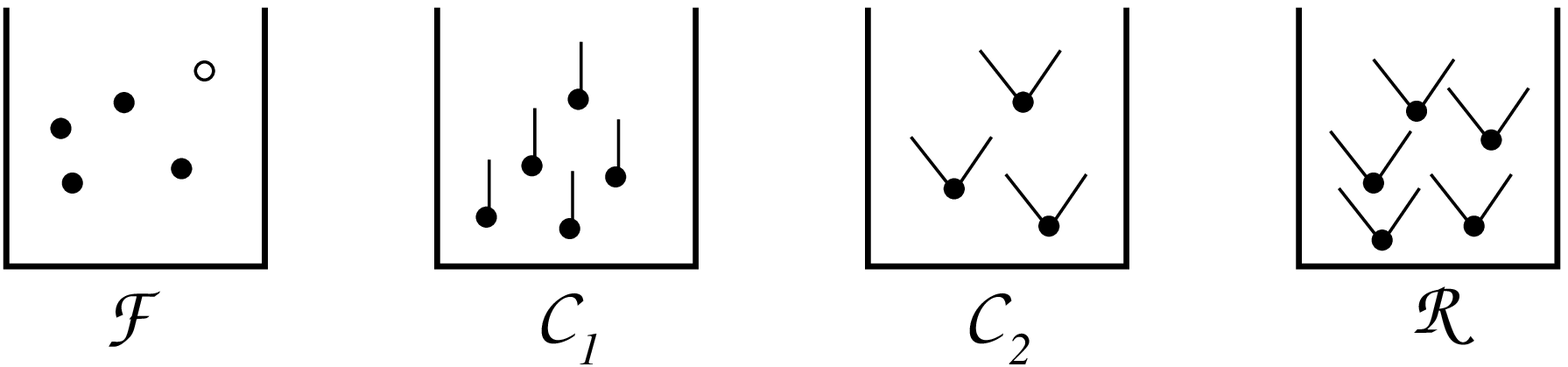}}\\
\parbox{1cm}{(2)}\parbox{7cm}{\includegraphics[width=7cm]{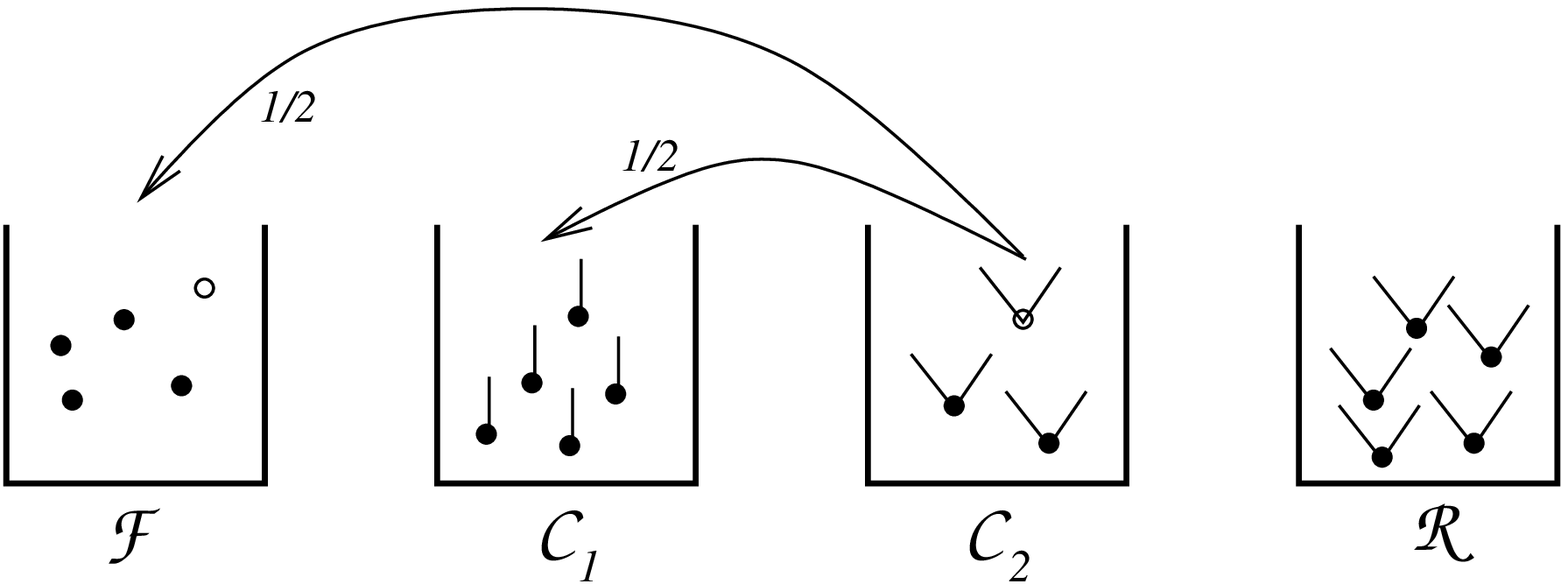}}\\
\parbox{1cm}{(3)}\parbox{7cm}{\includegraphics[width=7cm]{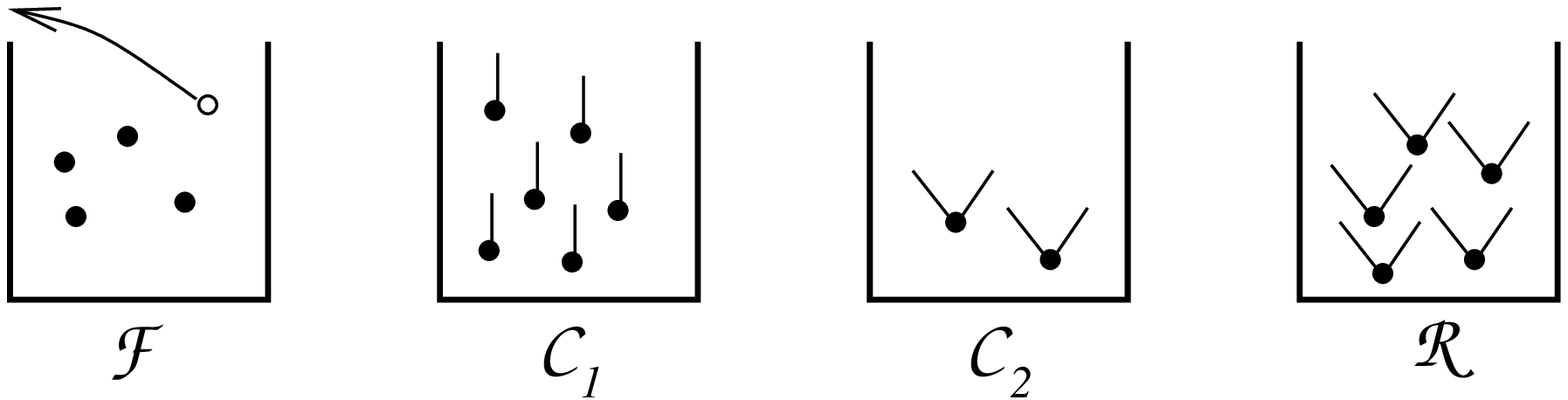}}\\
\parbox{1cm}{(4)}\parbox{7cm}{\includegraphics[width=7cm]{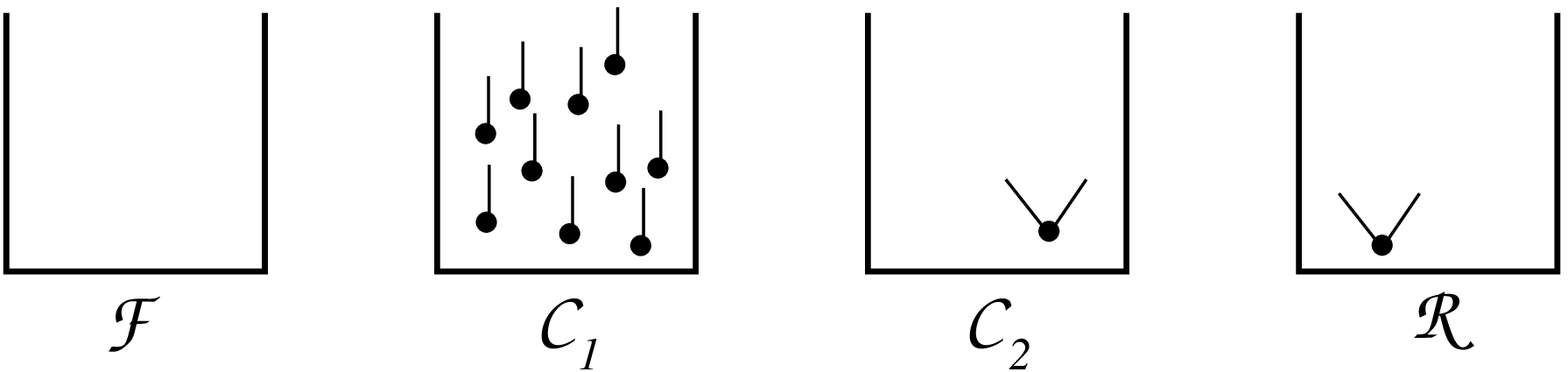}}\\
\caption{Illustration of the freezing process. (1) Initially, a frozen node is chosen (marked in white), (2) then it is determined to which node(s) this is an input and the effect on those nodes is determined. (3) Then, the selected frozen node is removed. (4) The last picture sketches the final state, where all frozen nodes have been removed and most remaining nodes have 1 nonfrozen input.}
\end{center}
\label{fig10}
\end{figure}

The number of nodes in the containers, $N$, can be used instead of the
time variable, since it decreases by one during each step. The
equations for $N_r$ and $N_{c_2}$ can then be solved by going from a difference
equation to a differential equation,
$$\frac {\Delta N_r}{\Delta N} \simeq \frac {d N_r}{d N} =  -
\frac{2N_r}{N}\, ,$$
which has the solution
\begin{equation}
N_r = \frac{\beta N^2}{N^{ini}}\, , \qquad N_{c_2} = \frac{\gamma N^2}{N^{ini}}\, , \label{Nr}
\end{equation}
where we have now denoted the total number of nodes with $N^{ini}$,
since the value of $N$ changes during the process.  
Similarly, we find if we neglect the noise term for a moment
\begin{eqnarray}
N_f &=& N(\delta-\beta)+ \frac{\beta N^2}{N^{ini}}\, , \nonumber\\
N_{c_1} &=&  N (\alpha+\gamma+2\beta) - 2  \frac{N^2(\beta+\gamma) }{N^{ini}}
\, . \label{det}
\end{eqnarray}
From this result, one can derive again the phase diagram, as we did by
using the annealed approximation: For $\delta < \beta$, i.e. if there
are more frozen than reversible update functions in the network, we
obtain $N_f=0$ at a nonzero value of $N$, and the number of nonfrozen
nodes is proportional to $N^{ini}$. We are in the chaotic phase. For
$\delta > \beta$, there exists no solution with $N_f=0$ and $N >
0$. The network is in the frozen phase.  For the critical networks
that we want to focus on, we have $\delta=\beta$, and the process
stops at $N_f = 1 = \beta {N^2}/{N^{ini}}$ if we neglect noise. This
means that $N = \sqrt{N^{ini}/\beta}$ at the end of the process.  The
number of nonfrozen nodes would scale with the square root of the
network size. This is not what is found in numerical studies of
sufficiently large networks.  We therefore must include the noise
term. Noise becomes important only after $N_f$ has become small, when
most nodes are found in container $\mathcal{C}_1$, and when the
variance of the noise has become unity, $\langle \xi^2\rangle =
1$. Inserting the solution for $N_r$ into the equation for $N_f$, we
obtain then
\begin{equation}
\frac{dN_f}{dN} = \frac{N_{f}}{N} +  \frac{\beta N}{N^{ini}}+\xi\label{langevin}
\end{equation}
with the step size $dN=1$. We want to transform this into a
Fokker-Planck-equation.  Let $P(N_f,N)$ be the probability that there
are $N_f$ nodes in container $\mathcal{F}$ at the moment where there
are $N$ nodes in total in the containers. This probability depends on
the initial node number $N_{ini}$, and on the parameter $\beta$. The
sum
$$\sum_{N_f=1}^\infty P(N_f,N) \simeq\int_0^\infty P(N_f,N) dN_f $$ is
the probability that the stochastic process is not yet finished,
i.e. the probability that $N_f$ has not yet reached the value 0 at the
moment where the total number of nodes in the containers has decreased
to the value $N$. This means that systems that have reached $N_f=0$
must be removed from the ensemble, and we therefore have to impose the
absorbing boundary condition $P(0,N)=0$. Exactly in the same way as
with calculation (\ref{FP}), we obtain then
\begin{equation}
-\frac{\partial P}{\partial N} = \frac{\partial}{\partial
    N_f} \left( \frac{N_{f}}{N} +  \frac{\beta
  N}{N^{ini}} \right)P + \frac 1 2 \frac{\partial^2 P}{\partial
   N_f^2}\, . \label{FP1}
\end{equation}
We introduce the variables 
\begin{equation}
x = \frac {N_f}{\sqrt{N}} \hbox{ and } y = \frac{N}{(N^{ini}/\beta)^{2/3}}\label{defxy}
\end{equation}
and the function $f(x,y)= (N^{ini}/\beta)^{1/3}P(N_f,N)$. We will see
in a moment that $f(x,y)$ does not depend explicitely on the parameters
$N^{ini}$ and $\beta$ with this definition. The Fokker-Planck equation
then becomes
\begin{equation}
y\frac{\partial f}{\partial y} +f+\left(\frac x 2 + y^{3/2}\right)
    \frac{\partial f}{\partial
    x} + \frac 1 2 \frac{\partial^2 f}{\partial
   x^2} = 0\, . \label{FP2}
\end{equation}
Let $W(N)$ denote the probability that $N$ nodes are left at the
moment where $N_f$ reaches the value zero. It is
\begin{eqnarray*}
W(N)  &=& \quad \int_0^\infty P(N_f,N)dN_f -  \int_0^\infty
P(N_f,N-1)dN_f \, .
\end{eqnarray*}
Consequently,
\begin{eqnarray}
W(N) &=& \frac{\partial}{\partial N}\int_0^\infty P(N_f,N)
dN_f\nonumber\\
&=& (N^{ini}/\beta)^{-1/3}\frac{\partial}{\partial N}\sqrt N \int_0^\infty f(x,y)dx
\nonumber\\
&=&  (N^{ini}/\beta)^{-2/3} \frac{\partial}{\partial y}\sqrt y \int_0^\infty f(x,y)dx
\nonumber\\
&\equiv&  (N^{ini}/\beta)^{-2/3}  G(y)\label{w}
\end{eqnarray}
with a scaling function $G(y)$. $W(N)$ must be a normalized function,
$$\int_0^\infty W(N)dN = \int_0^\infty G(y) dy = 1\, .$$ This
condition is independent of the parameters of the model, and therefore
$G(y)$ and $f(x,y)$ are independent of them, too, which justifies our
choice of the prefactor in the definition of $f(x,y)$.  The mean
number of nonfrozen nodes is therefore
\begin{equation}\bar N = \int_0^\infty N W(N)dN =  (N^{ini}/\beta)^{2/3}  \int_0^\infty G(y)
ydy\, ,\label{barN}
\end{equation}
which is proportional to $ (N^{ini}/\beta)^{2/3}$.  From Equations
(\ref{Nr}) and the corresponding equation for the $\mathcal{C}_2$-nodes we find
then that the number of nonfrozen nodes with two nonfrozen inputs is
proportional to $N^{1/3}$. This is a vanishing proportion of all
nonfrozen nodes. 

The nonfrozen nodes receive their (remaining) input from each other,
and we obtain the nonfrozen part of the network by randomly making the
remaining connections. If we neglect for a moment the second input of
those nonfrozen nodes that have two nonfrozen inputs, we obtain a
$K=1$ network. The number of relevant nodes must therefore be
proportional to the square root of number of nonfrozen nodes, i.e. it
is $N_{rel} \sim N^{1/3}$, and the number of relevant components is of
the order $\ln N^{1/3}$, with the largest component of the order of
$N^{2/3}$ nodes (including the trees). Adding the second input to the
nonfrozen nodes with two nonfrozen inputs does not change much: The
total number of relevant nodes that receive a second input is a
constant (since each of $\sim N^{1/3}$ relevant nodes receives a
second input with a probability proportional to $N^{-1/3}$). Only the
largest loops are likely to be affected, and therefore only the large
relevant components may have a structure that is more complex than a
simple loop.  Most nonfrozen nodes with two nonfrozen inputs sit in
the trees, as we have seen in Figure 5.3.  The mean number and
length of attractors can now be estimated in the following way: The
attractor number must be at least as large as the number of cycles on
the largest relevant loop, and therefore it increases exponentially
with the number of relevant nodes. The mean attractor length becomes
larger as for $K=1$ networks, since complex relevant components can
have attractors that comprise a large part of their state space, as
was shown in \cite{viktor:components}. Such components arise with a nonvanishing
probability, and they dominate therefore the mean attractor length,
which therefore increases now exponentially with the number of
relevant nodes.

The conclusions derived in the last paragraph can be made more
precise. Interested readers are referred to \cite{kaufmanandco:ourscaling}.

All these results are also valid for $K=2$ networks with only
canalyzing functions. As mentioned before, the frozen core of
canalyzing networks arises through self-freezing loops. The resulting
power laws are the same as for networks with constant functions, as was
shown in \cite{ute:canal}.

\subsection{Problems}
\begin{enumerate}
\item What is the number of attractors of the network shown in Figure
5.3 for all four cases where loop 1 and/or loop 2 are even/odd?
\item Assume there are 4 relevant nodes, one of them with two relevant inputs. List all topologically different possibilities for the relevant components.
\item Using Equation (\ref{w}), figure out how the probability that the entire network freezes depends on $N$. 
\end{enumerate}

\section{Networks with larger $K$}

Just as we did for $K=2$, we consider larger values of $K$ only for
those update rules that lead to fixed points of $b_t$ (i.e. of the
proportion of 1s), and therefore to a critical line separating a
frozen and a chaotic phase.

Let us first consider the frozen phase, where the sensitivity
$\lambda$ is smaller than 1. The probability that a certain node is
part of a relevant loop of size $l$ is for large $N$ obtained by the
following calculation: the node has $K$ inputs, which have again each
$K$ inputs, etc., so that there are $K^{l-1}$ nodes that might choose
the first node as one of its $K$ inputs, leading to a \emph{connection
loop}. The chosen node is therefore part of $K^l/N$ connection loops
of length $l$ on an average. The probability that a given connection
loop has no frozen connection is $(\lambda/K)^l$, and therefore the
mean number of relevant loops of size $l$ is $\lambda^l/l$. The mean
number of relevant nodes is then \begin{equation} \langle N_{rel}\rangle = \sum_l
\lambda^l = \frac{\lambda}{1-\lambda}. \label{relfrozen} \end{equation} This is
the same result as (\ref{relfrozen1}), which we derived for $K=1$.
The mean number of nonrelevant nodes to which a change of the state of
a relevant node propagates is given by the same sum, since in each
step the change propagates on an average to $\lambda$ nodes. By adding
the numbers of relevant and nonrelevant nonfrozen nodes, we
therefore obtain again a mean number of $\lambda/(1-\lambda)^2$
nonfrozen nodes, just as in the case $K=1$. We conclude that the
frozen phases of all RBNs are very similar.

Now we consider critical networks with $K>2$. The number of nonfrozen
nodes scales again as $N^{2/3}$ and the number of relevant nodes as
$N^{1/3}$. The number of nonfrozen nodes with $k$ nonfrozen inputs
scales with $N$ as $N^{(3-k)/3}$. These results are obtained by
generalizing the procedure used in the previous section for
determining the frozen core \cite{tamara}. By repeating the considerations of the
previous paragraph with the value $\lambda=1$, we find that in all
critical networks the mean number of relevant loops of size $l$ is
$1/l$ -- as long as $l$ is smaller than a cutoff, the value of which
depends on $N$. For $K=1$ the cutoff is at $\sqrt{N}$, for $K=2$, it
is at $N^{1/3}$, and this value does not change for larger $K$.  There
exists a nice phenomenological argument to derive the scaling $\sim
N^{2/3}$ of the number of nonfrozen nodes \cite{leerieger}: The number
of nonfrozen nodes should scale in the same way as the size of the
largest perturbation, since the largest perturbation affects all nodes
on the largest nonfrozen component. The cutoff $s_{max}$ in the size
of perturbations (see Equation (\ref{32})) is given by the condition
that $n(s_{max}) \sim 1/N$. Perturbations larger than this size occur
only rarely in networks of size $N$, since $n(s)$ is the probability
that a perturbation of one specific node (out of $N$ the nodes)
affects $s$ nodes in total. Using Equation (\ref{32}), we therefore
obtain \begin{equation} s_{max} \sim N^{2/3} \, .\end{equation} This argument does not work
for $K=1$, where we have obtained $s_{max} \sim N$ in section
\ref{k1}. The reason is that critical networks with $K=1$ have no
frozen core, but every node that receives its input from a perturbed
node will also be perturbed.

As far as the chaotic phase is concerned, there are good reasons to
assume that it displays similar features for all $K$. We have
explicitely considered the case $K=N$. Numerical studies show that the
basin entropy approaches a constant with increasing $K$ also when the
value of $K$ is fixed \cite{krawitz}. When $\lambda$ is close to 1,
there is a frozen core that comprises a considerable part of the
network. We can expect that the nonfrozen part has a state space
structure similar to that of the $K=N$ networks.

\section{Outlook}
\label{outlook}

There are many possibilities of how to go beyond RBNs with synchronous
update. In this last section, we will briefly discuss some of these
directions.

\subsection{Noise}

Synchronous update is unrealistic since networks do not usually have a
central pacemaker that tells all nodes when to perform the next
update. Asynchronous update can be done either deterministically by
assigning to each node an update time interval and an initial phase
(i.e. the time until the first update), or stochastically by assigning
to each node a time-dependent probability for being updated. We focus
here on stochastic update, since all physical systems contain some
degree of noise. In particular, noise is ubiquitous in gene regulatory
networks \cite{arkin97}.  Boolean networks with stochastic update are
for instance investigated in \cite{klemm05,greil:asyn}. The frozen
core obviously remains a frozen core under stochastic update, and the
relevant nodes remain relevant nodes.  The most fundamental change
that occurs when one switches from deterministic to stochastic update
is that there is now in general more than one successor to a
state. The set of recurrent states comprises those states that can
reoccur infinitely often after they have occurred for the first
time. However, if there is a path in state space from each state to a
fixed point or to a cycle that has only one successor for each state,
the network behaves deterministically for large times, in spite of the
stochastic update. This occurs in networks where all relevant nodes
sit on loops: an even loop has two fixed points, and an odd loop has
an attractor cycle of length $2l$, where each state has only one
successor in state space (apart from itself). If the number of
relevant loops increases logarithmically with system size $N$, the
number of attractors then increases as a power law of $N$. This means
that critical $K=1$ networks with asynchronous update have attractor
numbers that increases as a power law with system size. In
\cite{greil:asyn} it is argued that in critical $K=2$ networks, where
not all relevant components are simple loops, the attractor number is
still a power law in $N$.

The situation becomes different when the noise does not only affect
the update time but also the update function. Then the output of a
node can deviate from the value prescribed by the update function with
a probability that depends on the strength of the noise. The
interesting question to address in this context is whether the
networks remain in the neighborhood of one attractor (which can be
tested by evaluating the return probability after switching off the
noise), or whether they move through large regions of state
space. Investigations of networks with such a type of noise can be found
in \cite{shmu02,qu06}.

\subsection{Scale-free networks and other realistic network structures}

Real networks do not have a fixed number of inputs per node, but do
often have a power-law distribution in the number of inputs or the
number of outputs \cite{albert:review}. Boolean dynamics on such networks has been studied \cite{aldana03}, however, how this affects the power laws
in critical networks, is only partially known \cite{leerieger}.

There are many more characteristics of real networks that are not
found in random network topologies, such as clustering, modularity, or
scale invariance. The effect of all these features on the network
dynamics is not yet sufficiently explored.

\subsection{External inputs}

Real networks usually have some nodes that respond to external
inputs. Such an external input to a node can be modelled by switching
the constant function from 1 to 0 or vice versa. The set of nodes that
cannot be controlled in this way is called the computational
core. Networks with a higher proportion of $C_2$ functions tend to
have a larger computational core, since the $C_2$ functions can
mutually fix or control each other. Investigations of this type can be
found in \cite{correale06}.

\subsection{Evolution of Boolean networks}

Ensembles of networks that are completely different from the random
ensembles studied in this review can be generated by evolving networks
using some rule for mutations and for the network ``fitness''. For
instance, by selecting for robustness of the attractors under noise,
one obtains networks with short attractors that have large basins of
attraction, but that do not necessarily have a large frozen core
\cite{BS1,BS2,SD}. 

In another class of evolutionary models, fitness is not assigned to
the entire network, but links or functions are changed if they are
associated with nodes that do not show the ``desired'' behavior, for
instance if they are mostly frozen (or active), or if they behave most
of the time like the majority of other nodes \cite{bassler0,bassler1,bassler2}.

\subsection{Beyond the Boolean approximation}

There exist several examples of real networks, where the essential
dynamical steps can be recovered when using simple Boolean dynamics.
If a sequence of states shall be repeatable and stable, and if each
state is well enough approximated by an ``on''-''off'' description for
each node, Boolean dynamics should be a good approximation. However,
wherever the degree of activity of the nodes is important, the Boolean
approximation is not sufficent. This is the case for functions such as
continuous regulation or stochastic switiching or signal
amplification. Clearly, in those cases a modelling is needed that
works with continuous update functions or rate equations based on
concentrations of molecules. The different types of network modelling
are reviewed for instance in \cite{jongreview}.

\section*{Acknowledgement}
Many results reported in this article were obtained in collaboration
with former and present members of my group, Viktor Kaufman, Tamara
Mihalev, Florian Greil, Agnes Szejka.
I want to thank all those who read
earlier versions of this article and made suggestions for
improvements: Tamara Mihalev, Florian Greil, Agnes Szejka, Christoph
Hamer, Carsten Marr, Christoph Fretter.

\end{document}